## MEDICAL SCIENCES

# Induction in myeloid leukemic cells of genes that are expressed in different normal tissues

Joseph Lotem*, Hila Benjamin*†, Dvir Netaneli†, Eytan Domany† and Leo Sachs*¶

Departments of Molecular Genetics* and Physics of Complex Systems†, Weizmann Institute of Science, Rehovot 76100, Israel

*Contributed by Leo Sachs*

¶ To whom correspondence should be addressed:
Professor Leo Sachs, Department of Molecular Genetics, Weizmann Institute of Science, Rehovot 76100, Israel; tel. 972-8-934-4068; Fax. 972-8-934-4108;
E-mail: leo.sachs@weizmann.ac.il

Number of text pages, 19; Number of Figs, 3; Number of Tables, 1; Number of words in abstract, 241; Total number of characters in paper, 46586

Abbreviations: IL-6, interleukin 6; SPC, super paramagnetic clustering; TG, thapsigargin.

**Using DNA microarray and cluster analysis of expressed genes in a cloned line (M1-t-p53) of myeloid leukemic cells, we have analyzed the expression of genes that are preferentially expressed in different normal tissues. Clustering of 547 highly expressed genes in these leukemic cells showed 38 genes preferentially expressed in normal hematopoietic tissues and 122 other genes preferentially expressed in different normal non-hematopoietic tissues including neuronal tissues, muscle, liver and testis. We have also analyzed the genes whose expression in the leukemic cells changed after activation of wild-type p53 and treatment with the cytokine interleukin 6 (IL-6) or the calcium mobilizer thapsigargin (TG). Out of 620 such genes in the leukemic cells that were differentially expressed in normal tissues, clustering showed 80 genes that were preferentially expressed in hematopoietic tissues and 132 genes in different normal non-hematopietic tissues that also included neuronal tissues, muscle, liver and testis. Activation of p53 and treatment with IL-6 or TG induced different changes in the genes preferentially expressed in these normal tissues. These myeloid leukemic cells thus express genes that are expressed in normal non-hematopoietic tissues, and various treatments can reprogram these cells to induce other such non-hematopoietic genes. The results indicate that these leukemic cells share with normal hematopoietic stem cells the plasticity of differentiation to different cell types. It is suggested that this reprogramming to induce in malignant cells genes that are expressed in different normal tissues may be of clinical value in therapy.**

Normal hematopoietic stem cells express not only hematopoietic tissue-specific genes, but also genes expressed in various other normal tissues (1-3). In addition to the ability for self renewal, normal hematopoietic stem cells thus have transcription accessibility for genes expressed in non-hematopoietic tissues and this can explain the plasticity of these normal stem cells (4-6). Studies on the chromosomes (7-10) and other properties (11, 12) of cancer cells have shown that cancers also have self renewing stem cells (7-12). It was therefore of interest to determine whether cancer stem cells have the ability to express genes that are expressed in different normal tissues and how expression of such genes can be regulated under different conditions. To address this question we have used a cloned line of mouse myeloid leukemic cells (M1-t-p53) (4, 13, 14) that contains a temperature sensitive mutant p53 (V135A) protein which changes from a mutant form at 37ºC to wild-type form at 32ºC (15). Like the parental M1 myeloid leukemia cells (16-19), M1-t-p53 cells can be induced to undergo myeloid cell differentiation by the cytokine IL-6. When transferred from 37ºC to 32ºC, activation of wild-type p53 in M1-t-p53 cells induces apoptosis that can be effectively inhibited by IL-6 and the calcium mobilizer thapsigargin (TG) (13, 14, 18, 20).

We have previously used these cells for microarray analysis of changes in gene expression after activation of p53 in the absence or presence of IL-6 or TG (20). This analysis showed that p53-induced apoptosis can be inhibited without affecting expression of p53 regulated genes. IL-6 and TG inhibited p53 induced apoptosis by different pathways of gene expression, and IL-6 and TG induced expression of different hematopoietic differentiation-associated genes (20). DNA microarray analysis of gene expression in a panel of 45 normal mouse tissues has shown that about 80% of the genes are differentially expressed and that there are specific gene expression profiles in different normal tissues (21). We have now compared the gene expression in M1-t-p53 cells to the profile of gene expression in this panel of 45 normal mouse tissues. Our results indicate that these myeloid leukemic cells express genes that are preferentially expressed in different normal non-hematopoietic tissues, and that activation of p53 and treatment with IL-6 or TG can induce expression of other such non-hematopoietic genes. These leukemic cells thus share with normal hematopoietic stem cells the capacity to express genes that are expressed in different normal non-hematopoietic tissues.

## Materials and Methods

**Cells.** We have used M1-t-p53 mouse myeloid leukemic cells that express a temperature-sensitive p53 protein that changes from a mutant to wild-type form after transfer from 37ºC to 32ºC (15) and induces apoptosis at 32ºC but not at 37ºC (4, 13, 14, 18, 20). The cells were cultured and treated with IL-6 or TG as described (20). Growth curves at 37ºC and the formation of colonies from single cells show that M1-t-p53 and the parental M1 leukemic cells can self-renew (22).

**Analysis of tissue-specific gene expression profiles.** Two DNA microarray datasets were used: in the first dataset, mRNA expression levels of genes in M1-t-p53 mouse myeloid leukemic cells (20) were measured using the Affymetrix U74Av2 array, including 12,488 probe sets. In the second dataset, mRNA expression profiles of genes in 45 normal mouse tissues (21) were examined using the Affymetrix U74A array, including 12,588 probe sets. There were 9,977 probe sets in common between the two microarrays. For both datasets, the expression value for each gene was determined using the MicrroArray Suite version 5.0 (MAS 5.0) software (23) with default parameters. Gene expression values <20 were adjusted to 20 to eliminate noise from the data and all values were then $\log_{10}$ transformed.

To determine the normal expression profiles of genes that are expressed in untreated M1-t-p53 myeloid leukemic cells, we selected the most highly expressed genes in these cells, the 85$^{th}$ percentile, showing intensity values above 314 Affymetrix fluorescence units. We then checked the expression profiles of these genes in normal mouse tissues (21). The selected genes were filtered according to their expression in the 45 normal mouse tissues, to remove those genes that show a similar expression level in all 45 tissues. We used two criteria to filter the genes: **1.** Genes whose expression showed a high standard deviation in the different mouse tissues, $\geq$0.3 of their $\log_{10}$ transformed expression values; **2.** Genes whose expression level range was at least 1, in $\log_{10}$ transformed values, and whose expression in at least one tissue was $\geq$2 standard deviations below or above the mean. The genes that passed either of these two criteria, or both, were clustered according to their expression in the different mouse tissues, using a new, very efficient version (Barad, O., M.Sc. thesis, Weizmann Institute of Science, 2003;http://www.weizmann.ac.il/physics/complex/compphys/group_papers.htm#omer_b) of the Super-Paramagnetic clustering (SPC) algorithm (24). Before each clustering operation, the $\log_{10}$ transformed expression values as measured over all experiments were first centered and normalized such that the mean of its component is set to 0, and the sum of squares is set to 1 as described (25).

We have previously found 1786 genes whose expression changed after activation of wild-type p53 in these leukemic cells in the absence or presence of IL-6 (20). Because the data sets of M1-t-p53 cells and the normal mouse tissues were obtained using slightly different microarrays, only 1497 of the above 1786 M1-t-p53 genes could be compared to the data from mouse tissues. To search for tissue-specific expression, the genes with low variability of expression between tissues were filtered out of these 1497 genes as described above, and the selected genes were clustered, using SPC, according to their expression in the different mouse tissues. We have also carried out the same procedure of filtering and SPC clustering with 275 additional genes that were up-regulated >2-fold in at least 2 time points in M1-t-p53 cells after p53 activation in the absence or presence of TG.

## Results

**Initial expression of non-hematopoietic normal tissue-specific genes in M1-t-p53 myeloid leukemic cells.** To search for initially expressed genes in M1-t-p53 cells that are preferentially expressed in normal non-hematopoietic tissues, we compared a selected group of highly expressed genes in M1-t-p53 cells (20) to their expression profile in 45 normal mouse tissues (21). Out of the 12488 genes represented in the microarray, 1857 genes expressed in the M1-t-p53 cells at levels >314 Affymetrix fluorescence units, 85$^{th}$ percentile, were selected and 1613 of these genes were also present in the data set of Su et al. (21). These 1613 genes were then filtered to remove those genes that showed low variability of expression level in the 45 normal tissues, as described in Materials and Methods. This filtration detected 547 differentially expressed genes that were clustered according to their relative expression in different normal tissues. This clustering showed 38 genes in clusters that are preferentially expressed in normal hematopoietic tissues including bone and bone marrow (c1) or bone, bone marrow, lymph node, spleen and thymus (c2 and c3) (Fig. 1). In addition, M1-t-p53 cells also expressed 122 genes (22.3%) in clusters that are preferentially expressed in various non-hematopoietic tissues including neuronal tissues such as brain, dorsal root ganglia, spinal cord and some other neuronal tissues (cluster b, 58 genes), skeletal muscle, heart and brown fat (cluster a, 40 genes), liver, kidney and gall bladder (clusters d1 and d2, 13 genes) and testis (cluster e, 11 genes) (Fig. 1) (see the list of genes in Table 2, which is published as supporting information on the PNAS web site). We tested for dependence of our findings on the (arbitrary) choice of using the upper 15% of expression values, and found that the clusters described above were robust against varying the value of the threshold. Similar clusters were obtained when we selected for the analysis genes expressed in M1-t-p53 cells above the $80^{th}$ or $90^{th}$ percentiles, >219 or >479 Affymetrix fluorescence units, respectively. These results indicate that like normal hematopoietic stem cells, these leukemic cells express genes that are expressed in multiple normal non-hematopoietic tissues.

**Regulation of expression of non-hematopoietic normal tissue-specific genes in myeloid leukemic cells by wild-type p53, IL-6 and TG.** We have previously shown that activation of p53 in the absence or presence of IL-6 in M1-t-p53 cells induces changes in expression of 1786 genes (20). To determine whether these 1786 genes include genes that are preferentially expressed in normal non-hematopoietic tissues, we first filtered out those genes that showed a low variability in expression levels in the 45 normal tissues, as described in Materials and Methods. This filtration process detected 578 genes that showed differential expression in different tissues. Clustering these 578 genes according

to expression in different tissues identified 6 major clusters showing different patterns of tissue-specific gene expression (Fig. 2) (see the list of genes in these clusters and their regulation by activated p53 and IL-6 in Table 3, which is published as supporting information on the PNAS web site). The largest cluster (c) contained 68 genes preferentially expressed in hematopoietic tissues. This cluster can be sub-divided into two clusters, one (c1) preferentially expressed in bone marrow and bone and the other (c2) expressed in bone marrow, bone, lymph nodes, spleen, thymus, and also in trachea and adipose tissues (Fig. 2). The other clusters of genes expressed in these leukemic cells contained genes whose profiles showed preferential expression in different normal non-hematopoietic tissues. These clusters include genes preferentially expressed in neuronal tissues (cluster f, 39 genes), testis (cluster a, 37 genes), skeletal muscle, heart and brown fat (cluster b, 29 genes), liver, kidney and gall bladder (cluster d, 24 genes) and digits and epidermis (cluster e, 7 genes) (Fig. 2).

We also carried out a similar filtration of the 275 genes selected as described in Materials and Methods, whose expression was up-regulated in the myeloid leukemic cells by p53 in the absence or presence of TG. This filtration gave 116 up-regulated genes that were differentially expressed in normal tissues, which were clustered as above. Here the analysis also identified several clusters of genes showing different expression profiles in normal tissues (Fig. 3) (see the list of genes in these clusters and their regulation by activated p53 or TG in Table 4, which is published as supporting information on the PNAS web site). These clusters included genes preferentially expressed in hematopoietic tissues (clusters b and c, 21 genes), neuronal tissues (cluster a, 16 genes), skeletal muscle, heart and brown fat (cluster d, 7 genes), testis (cluster e, 4 genes) and liver and gall bladder (cluster f, 4 genes) (Fig. 3).

Comparison of the lists of 578 and 116 genes that were analyzed in the last two experiments showed 74 common genes. Thus, out of the 578+116-74=620 genes whose expression was changed in M1-t-p53 myeloid leukemic cells by p53, IL-6 or TG and which showed differential expression in different normal mouse tissues, 21.3% (132 genes) were preferentially expressed in non-hematopoietic tissues. Out of these 132 non-hematopoietic genes, p53 up-regulated expression of 59 genes, IL-6 up-regulated 10 genes and TG up-regulated 14 genes. Examples of genes up-regulated by p53, IL-6 and TG are given in Table 1. p53 also down-regulated expression of 32 non-hematopoietic genes and IL-6 down-regulated 17 such genes (see the lists of genes and their regulation by activated p53, IL-6 or TG in Tables 3 and 4, which are published as supporting information on the PNAS web site). Comparison of the lists of non-hematopoietic genes expressed in the leukemic cells before and after treatment with p53, IL-6 or TG has shown that only 9 of the above 132 non-hematopoietic tissue-specific genes were among the 122 non-hematopoietic genes that are initially expressed in M1-t-p53 cells. Thus, in addition to the 122 non-hematopoietic genes that are initially highly expressed in these leukemic cells, p53, IL-6 and TG could reprogram these cells to change expression of other genes that show preferential expression in normal non-hematopoietic tissues.

## Discussion

The ability of normal somatic stem cells to self-renew provides the body with the necessary cell reservoir to replenish cells that die. In addition to this capacity for self-renewal, normal hematopoietic stem cells (4-6) and normal stem cells from other sources (26) have a plasticity for differentiation to different cell types. The plasticity for differentiation of normal hematopoietic stem cells can be explained by the transcription accessibility in these cells for genes that are normally expressed in non-hematopoietic tissues (1-3). Since cancers also have stem cells that can self-renew (7-12), we asked whether leukemic cells, like their normal hematopoietic stem cell counterparts, also display transcription accessibility for genes that are normally expressed in different non-hematopoietic tissues. We have addressed this question using a whole genome expression approach and compared gene expression in the cloned mouse myeloid leukemic cell line M1-t-p53 (20) and in 45 normal mouse tissues (21). Study of the chromosomes of M1 myeloid leukemic clones have shown that more than 90% of the cells have the same chromosome number and all the cells in a clone have the same chromosome abnormalities (10). The growth curve of M1-t-p53 cells has shown that these cells have a high rate of cell multiplication and 50% of the cells formed colonies from single cells when cloned in agar (22). The M1-t-p53 leukemia thus has a high frequency of self-renewing stem cells.

Analysis of gene expression in mouse and human normal tissues has shown that about 80% of the genes show a differential expression in different tissues, which can be used to identify tissue specific gene expression profiles (21). We searched for genes that are initially highly expressed in M1-t-p53 leukemic cells and are preferentially expressed in normal non-hematopoietic tissues. We found 122 such genes in clusters whose expression profile showed preferential expression in different normal non-hematopoietic tissues including neuronal tissues, muscle, liver and testis. These results indicate that like normal hematopoietic stem cells, these leukemic cells express genes that are expressed in non-hematopoietic normal tissues.

The use of M1-t-p53 myeloid leukemic cells enabled us to determine the regulation of expression of non-hematopoietic genes under different conditions including after activation of wild-type p53 and treatment with IL-6 or TG. We found 620 genes whose expression in the leukemic cells changed under these conditions and which were differentially expressed in different normal tissues. Out of these 620 genes, 132 genes were preferentially expressed in different normal non-hematopoietic tissues that also included neuronal tissues, muscle, liver and testis. Out of these 132 genes, 59 genes were up-regulated in the leukemic cells by p53, 10 by IL-6 and 14 by TG. Only 9 out of these 132 non-hematopoietic tissue-specific genes, whose expression was regulated by p53, IL-6 or TG, were among the 122 non-hematopoietic tissue-specific genes initially expressed in M1-t-p53 cells. Our results indicate that in addition to the 122 genes that are preferentially expressed in non-hematopoietic tissues and were initially expressed in the leukemic cells, p53, IL-6 and TG could reprogram these cells to change expression of other genes that are preferentially expressed in normal non-hematopoietic tissues.

The major mechanism for the transcription accessibility in the leukemic cells for genes normally expressed in non-hematopoietic tissues, presumably involves changes in DNA methylation and other epigenetic changes (19, 27-31). Induction of DNA demethylation and inhibition of histone deacetylase activity by deoxycytidine and trichostatin A, respectively, can up-regulate expression of many genes that are silenced in cancer cells (32). It will be interesting to determine whether genes that are up-regulated by such compounds in cancer cells also include genes that are preferentially expressed in normal tissues other than the tissue in which the cancer originated. Alternative splicing may also play a role in expression of non-hematopoietic genes in the leukemic cells. One of the genes that is preferentially expressed in a non-hematopoietic tissue and is up-regulated by p53 in the M1-t-p53 leukemic cells (20) is the gene Ak1 (Adenylate kinase 1). Using the same temperature-sensitive p53 (V135A) that we used, it was shown that an alternatively spliced Ak1 transcript can be activated by wild-type p53 (33). The Mel1s and Hox11 genes that are highly expressed in some leukemias but not in normal blood cells are also alternatively spliced transcripts (34, 35). Alternatively spliced transcripts of apparently tissue-specific genes may thus be expressed in cells of another lineage.

Our results indicate that the M1 leukemic cells share with normal hematopoietic stem cells the capacity to express genes that are preferentially expressed in non-hematopoietic normal tissues. These results and the finding that under different conditions other non-hematopoietic genes were induced in these leukemic cells indicate that the leukemic cells share with normal hematopoietic stem cells the plasticity for differentiation to different cell types. Cancer therapy includes the induction of differentiation in cancer cells (16, 17, 19, 36). It will be of interest to determine to what extent different types of cancer have differences in the plasticity of gene expression for different normal tissues, and how this will affect the growth and metastasis of the cancer and its clinical response to therapy.

This work was supported by the Benoziyo Institute of Molecular Medicine, the Dolfi and Lola Ebner Center for Biomedical Research, Dr. and Mrs. Leslie Bernstein, Mrs. Bernice Gershenson, the Israel Academy of Sciences (ISF), the Minerva Foundation and the Ridgefield Foundation.

**Table 1. Examples of up-regulation in leukemic cells by p53, IL-6 or TG of genes that are expressed in different normal tissues***

| Tissues | p53 | | | IL-6 | | | TG | | |
|---|---|---|---|---|---|---|---|---|---|
| Hematopoietic | 22 genes | | | 13 genes | | | 8 genes | | |
| | Gtse1 | AJ222580 | 11.5 | Il4ra | M27960 | 15.0 | Adcy | U12919 | 3.3 |
| | Tap1 | U60020 | 5.7 | Cd53 | X97227 | 4.9 | Hcph | M68902 | 2.8 |
| Neuronal | 26 genes | | | 2 genes | | | 8 genes | | |
| | Ptpro | U37465 | 6.2 | Egr2 | M24377 | 3.6 | Egr2 | M24377 | 9.6 |
| | Pmm1 | AF007267 | 5.9 | Snx10 | AI746846 | 3.4 | Serpini1 | AJ001700 | 7.8 |
| Muscle | 11 genes | | | 1 gene | | | 2 genes | | |
| | Ak1 | AJ010108 | 35.6 | Sh3glb1 | AW125190 | 3.8 | Kpna1 | U20619 | 2.7 |
| | Lpin1 | AI846934 | 12.6 | | | | Mlf1 | AF100171 | 2.0 |
| Liver | 8 genes | | | 3 genes | | | 3 genes | | |
| | Ccsd | AI839702 | 4.2 | Sc5d | AB016248 | 2.7 | Lifr | D17444 | 4.4 |
| | Lypla | AA840463 | 3.3 | Lifr | D17444 | 2.5 | Cp | U49430 | 3.3 |
| Testis | 9 genes | | | 3 genes | | | 1 gene | | |
| | Tcte3 | U21673 | 12.0 | Taf9 | AI842969 | 2.2 | Xmr | X72697 | 2.6 |
| | Ppm1b | D45859 | 4.6 | Nphp1 | AJ243223 | 2.5 | | | |

Legend to Table 1

* The genes shown in this Table are examples of up-regulated genes in M1-t-p53 leukemic cells in clusters of genes that are preferentially expressed in normal hematopoietic and non-hematopoietic tissues. Each gene is shown in the Table by its name, GeneBank accession number and fold-up-regulation by p53, IL-6 or TG. Fold up-regulation is from results described in Lotem et al. (20). All the genes regulated by p53, IL-6 and TG are shown in Tables 3 and 4, which are published as supporting information on the PNAS web site.

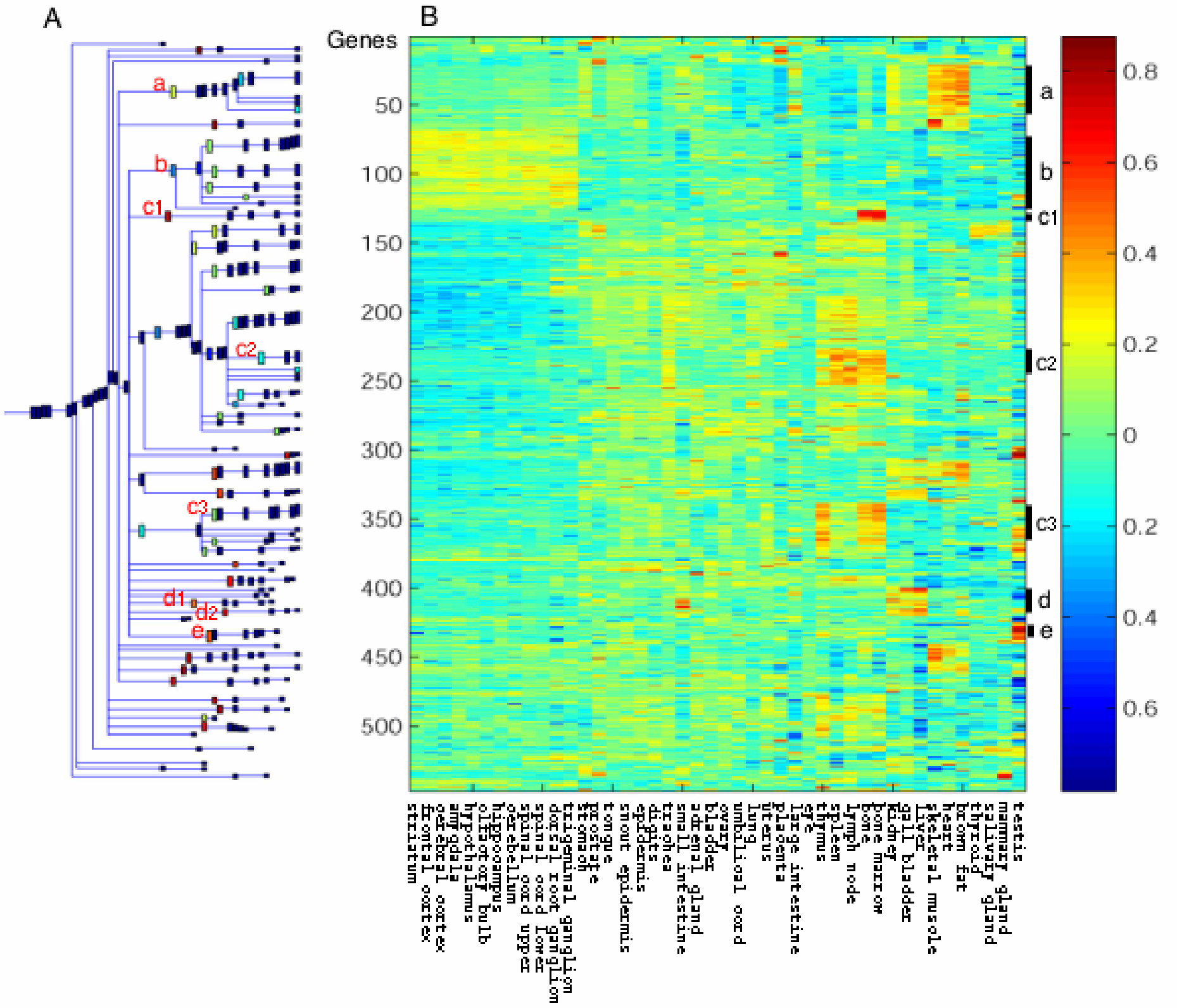

**Fig.1.** M1-t-p53 myeloid leukemic cells initially express genes that are preferentially expressed in different normal tissues. Highly expressed genes in the leukemic cells, which were differentially expressed in normal mouse tissues, were clustered according to their relative expression in normal mouse tissues using SPC, as described in Materials and Methods. A, dendrogram of the 547 differentially expressed genes that include clusters of at least 5 genes. Selected clusters showing genes with similar relative expression in different normal tissues are marked by the red letters a-e. B, Re-ordered expression matrix according to the clustering result, where each row represents a gene and each column represents a tissue. The clustering operation imposes a linear ordering of the data points and the rows and columns were ordered accordingly. The colors indicate the relative expression levels of the genes in 45 normal tissues according to the color code shown on the right side of the panel. The bars and letters on the right side of the panel represent the selected gene clusters a-e.

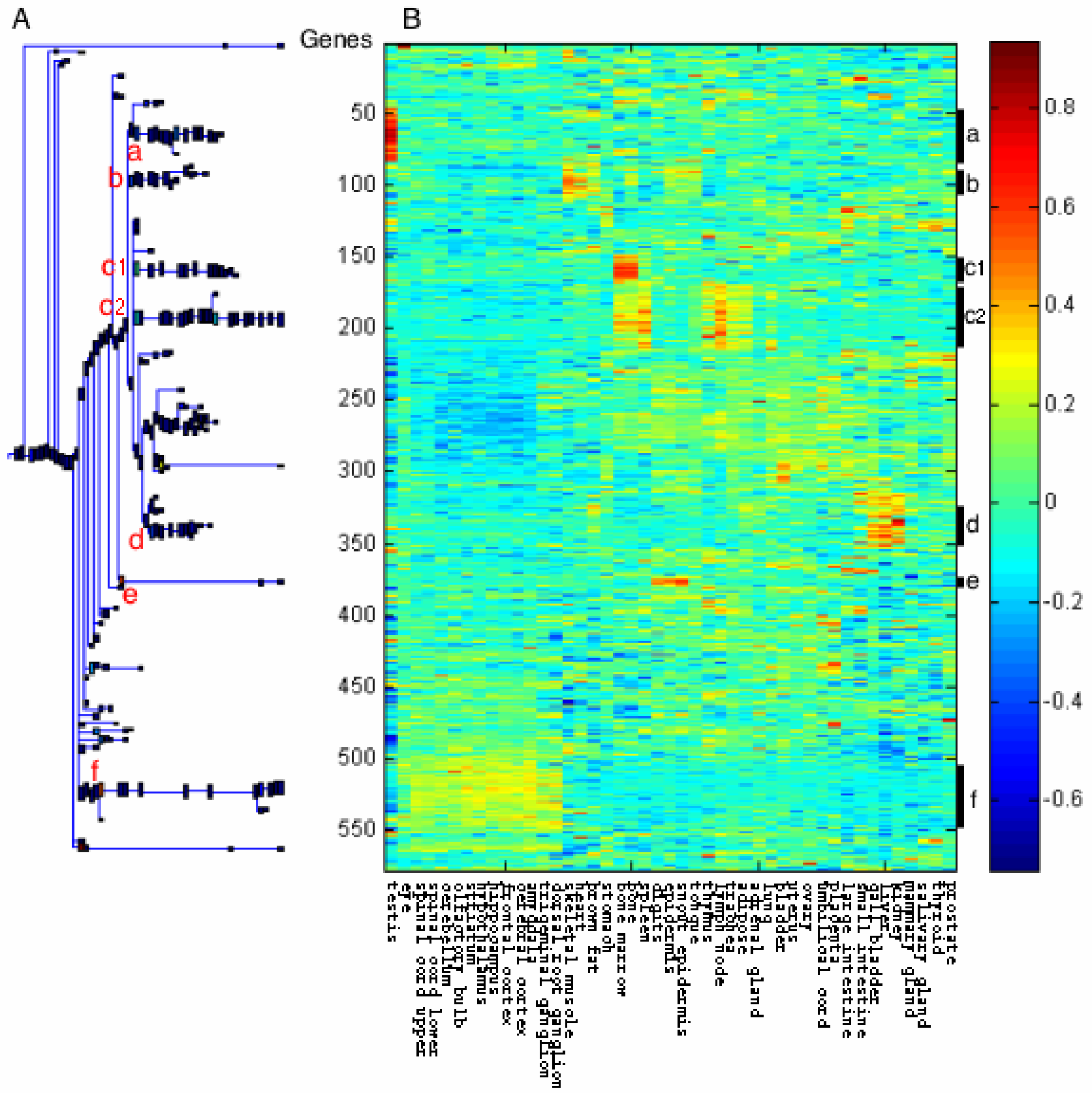

**Fig.2.** Regulation by wild-type p53 and IL-6 in M1-t-p53 myeloid leukemic cells of genes that are preferentially expressed in different normal tissues. Genes whose expression in the leukemic cells was changed after activation of wild-type p53 or IL-6 (20) and which were differentially expressed in normal mouse tissues, were clustered, using SPC, according to their relative expression in normal mouse tissues. Minimal cluster size shown is 5 genes per cluster. Clusters of genes showing preferential expression in different normal tissues are marked as in Fig. 1.

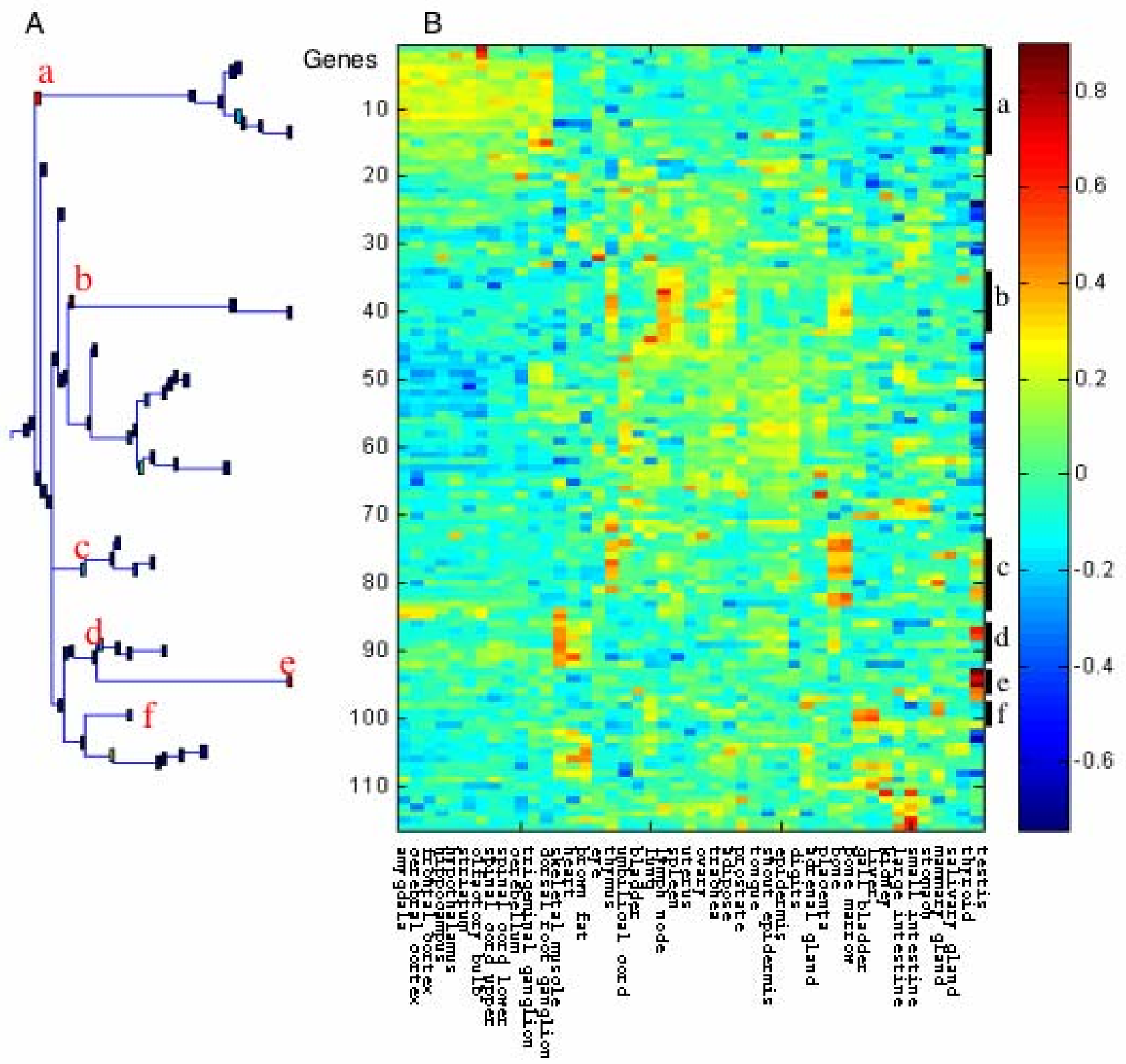

**Fig.3.** Regulation by wild-type p53 and TG in M1-t-p53 myeloid leukemic cells of genes that are preferentially expressed in different normal tissues. Genes whose expression in the leukemic cells was up-regulated >2-fold after activation of wild-type p53 or TG (20) and which were differentially expressed in normal mouse tissues, were clustered, using SPC, according to their relative expression in normal mouse tissues. Minimal cluster size shown is 4 genes per cluster. Clusters of genes showing preferential expression in different normal tissues are marked as in Figs. 1 and 2.